\documentclass[twocolumn]{revtex4}

\usepackage{graphicx}
\usepackage{dcolumn}
\usepackage{amssymb}
\usepackage{bm}
\usepackage{epsfig}
\usepackage{psfrag}
\usepackage{float}

\def\3dots{\:\raisebox{-0.5ex}{$\stackrel{\textstyle.}{:}$}\:}
\def\beq{\begin{equation}}
\def\eeq{\end{equation}}
\def\bea{\begin{eqnarray}}
\def\eea{\end{eqnarray}}
\def\correspondingauthors{\footnote{ratheevikram@gmail.com}}
\def\correspondingauthor{\footnote{dlb76@georgetown.edu}}
\def\correspondingauth{\footnote{urbachj@georgetown.edu}}

\begin{document}

\title{Dynamics and memory of boundary stresses  in discontinuous shear thickening suspensions during oscillatory shear}

\author{Vikram Rathee\correspondingauthors{}\footnote{Present Address: Okinawa Institute of Science and Technology Graduate University, Okinawa 904-0495, Japan}}

\affiliation{Department of Physics, Georgetown University, Washington, DC 20057; and Institute for Soft Matter Synthesis and Metrology, Georgetown University,
Washington, DC 20057  }

\author{Daniel L. Blair \correspondingauthor{}}
\affiliation{Department of Physics, Georgetown University, Washington, DC 20057; and Institute for Soft Matter Synthesis and Metrology, Georgetown University,
Washington, DC 20057 }

\author{Jeffrey S. Urbach \correspondingauth{}}
\affiliation{Department of Physics, Georgetown University, Washington, DC 20057; and Institute for Soft Matter Synthesis and Metrology, Georgetown University,
Washington, DC 20057
}

\date{\today}

\begin{abstract}

We report direct measurements of spatially resolved surface stresses of a dense suspension during large amplitude oscillatory shear (LAOS) in the discontinuous shear thickening  regime using boundary stress microscopy.  Consistent with previous studies,  bulk rheology shows a dramatic increase in the complex viscosity above a frequency-dependent critical strain.  We find that the viscosity increase is coincident with that appearance of large heterogeneous boundary stresses, indicative of the formation of transient solid-like phases (SLPs) on spatial scales large compared to the particle size. The critical strain for the appearance of SLPs is largely determined by the peak oscillatory stress, which depends on the peak shear rate and the frequency-dependent suspension viscosity.  The SLPs dissipate and reform on each cycle, with a spatial pattern that is highly variable at low frequencies but remarkably persistent at the highest frequency measured ($\omega = 10$ rad/sec).
\end{abstract}

\maketitle

\section{Introduction}
Dense colloidal and granular suspensions display a variety of non-Newtonian behaviors, including a dramatic increase in viscosity, $\eta$, above a material dependent critical shear stress (reviewed in \cite{Morris:2020aa}). When the particle volume fraction $\phi=V_{\rm{particle}}/V_{\rm{total}}$ is sufficiently high, suspensions can exhibit discontinuous shear thickening (DST), where the viscosity as a function of shear rate, measured during steady shear, increases abruptly by more than an order of magnitude.

While shear thickening can arise from the formation of particle clusters due to hydrodynamic forces \cite{Wagner:2009aa, Melrose:1996aa, Bossis:1989aa, Gurnon:2015aa}, solid contact and friction, together with the short-range repulsive force between particles, are thought to be necessary for DST \cite{Cates:2014aa, Seto:2013aa, Mari:2014aa, Mari:2015aa}. Wyart and Cates (WC) introduced a phenomenological model incorporating these elements, and a variety of results from experiments and simulations provide support for  this model and its extensions \cite{Morris:2020aa}, although many questions remain.  

One set of questions concerns the spatiotemporal dynamics of DST.  Suspensions subjected to constant shear stress or stain rate display a variety of complex  fluctuations, including temporal fluctuations in bulk viscosity, and  visual observations suggest some associated spatial heterogeneities \cite{Lootens:2003aa, Guy:2015aa, Saint:2018aa, Nayoung:2019aa}.  We recently reported direct measurements of spatially resolved surface stresses over the entire surface of a dense suspension during DST under steady shear using boundary stress microscopy (BSM, \cite{Rathee:2017aa, Arevalo:2015aa}). We found that large fluctuations in the bulk rheological response at the onset of DST are the result of localized transitions to a state with very high stress, consistent with a fully jammed solid that makes direct contact with the shearing boundaries \cite{Rathee:2020aa}.  Here, we report the application of BSM to a suspension undergoing DST during Large Amplitude Oscillatory Shear (LAOS).  

LAOS is a powerful tool for characterizing the nonlinear response of complex fluids and soft solids \cite{Hyun:2011aa}, but its application to shear thickening suspensions has been somewhat limited \cite{Morris:2020aa}.  Studies of athermal (granular) suspensions have revealed that a  strain amplitude $\gamma_o$ of order $0.1$ is required to initiate particle contact and the attendant nonlinear increase in viscosity \cite{Ness:2017aa}, and that a transition from reversible to chaotic particle trajectories is associated with yeilding of the jammed phase, again at strain amplitudes below unity \cite{Galloway:2020aa}.  Studies at higher strain amplitudes revealed that the appearance of strong shear thickening is primarily driven by the peak strain rate, rather than the peak strain \cite{Khandavalli:2015aa}.  LAOS in combination with rapid shear reversal or shear cessation has been used to determine how different contributions to the overall  stress evolve over an oscillation cycle \cite{Ong:2020aa}.  Using small angle x-ray scattering, Lee et al. \cite{Lee:2018aa} revealed that, under certain conditions, a two-step thickening process can occur, with a strain-controlled order-to-disorder transition leading to moderate shear thickening, followed by a strain rate or stress controlled transition to a dramatically increased viscosity.  
For the sample studied here, nearly monodisperse Brownian spheres at a concentration that produces DST, bulk rheology during LAOS shows two step thickening, consistent with \cite{Lee:2018aa}.  Using BSM, we show the the boundary stresses are uniform in the low viscosity and first thickened regime, but become strongly nonuniform coincident with the more dramatic shear thickening regime.  At low frequencies, regions of very high stress appear on each cycle at a shear rate that is comparable to the rate required to induce DST in steady shear, with spatiotemporal dynamics that are consistent with the picture derived from our BSM measurement under steady shear \cite{Rathee:2020aa}: localized jammed solid like phases (SLPs), spanning the rheometer gap, form above a critical shear rate, but are immediately fractured, producing two separate SLPs that propagate in opposite directions, one that remains in contact with the bottom boundary and the other remains in contact with the top. At higher frequencies the shear rate required for the onset of SLPs increases, an effect that is attributable to a frequency-dependent decrease in viscosity, likely arising from particle ordering. The  spatial pattern of the large stresses is highly variable at low frequencies but remarkably persist at high frequencies, indicating that the suspension retains memory of the configurations responsible for the SLPs throughout the oscillation cycle.

\section{Materials and methods}
Colloidal suspensions were formulated with silica spheres of radius $a$ = 0.75 $\mu$m (Angstorm, Inc.) suspended in a glycerol water mixture (0.8 glycerol volume fraction). The particle volume fraction $\phi$ was calculated from the mass used to prepare the suspension. Rheological measurements were performed on a stress-controlled rheometer (Anton Paar MCR 301) mounted on an inverted confocal (Leica SP5) microscope \cite{Dutta:2013aa} using a cone of diameter 25 mm. Elastic substrates were deposited on glass cover slides of 40 mm diameter by spin coating PDMS (Sylgard 184; Dow Corning) and a curing agent with elastic modulus $\sim$ 15 kPa \cite{Rathee:2020aa}.  Beads of 10 $\mu$m were covalantly attached to the PDMS and an area of 5.6 mm$^{2}$  was imaged, centered approximately 3 mm from the outer edge of the rheometer tool.  Deformation fields were determined with particle image velocimetry (PIV) in ImageJ \cite{Tseng:2012aa}. The surface stresses at the interface are calculated using an extended traction force technique and codes given in ref. \cite{Style:2014aa}.  Taking the component of the surface stress in the flow (velocity) direction, we obtain the  scalar field $\sigma_{BSM}(\vec{r},t)$, representing the spatiotemporally varying surface stress.  The overall accuracy of the BSM is limited by uncertainty in the modulus of the PDMS layer \cite{Rathee:2020aa}, and measurement noise arises from the resolution of the PIV imaging and residual vibrations in the imaging system.

Oscillatory shear measurements were performed by applying a sinusoidal shear strain, $\gamma(t) = \gamma_{o} \sin(\omega t)$, 
where $\gamma_{o}$ is the strain amplitude  and $\omega$ is the angular frequency. 
For a small strain amplitudes the stress response is given by  $\sigma(t) = \sigma_{o} \sin(\omega t + \delta)$, where $\delta$ is the phase lag.  At larger strains where the response is non-linear, higher harmonics contribute to the stress response which can be expressed as   $\sigma(t) = \sum \sigma_{n} \sin(n\omega t +\delta_{n})$. The elastic and loss modulii and the complex viscosity are calculated from the amplitude and phase of the fundamental, $G'=(\sigma_1/\gamma_o) \cos{\delta_1}$, $G''=(\sigma_1/\gamma_o) \sin{\delta_1}$, $\eta ^*= \sigma_1/(\omega \gamma_o)$.
The graphical representation of stress versus applied strain or strain rate is knows as a Lissajous figure.  During LAOS measurements, the stress vs. strain data reported by the rheometer represent averages over four consecutive cycles. 


\section{Large Amplitude Oscillatory Shear (LAOS) Rheology}

\begin{figure}
\includegraphics[width=0.45\textwidth]{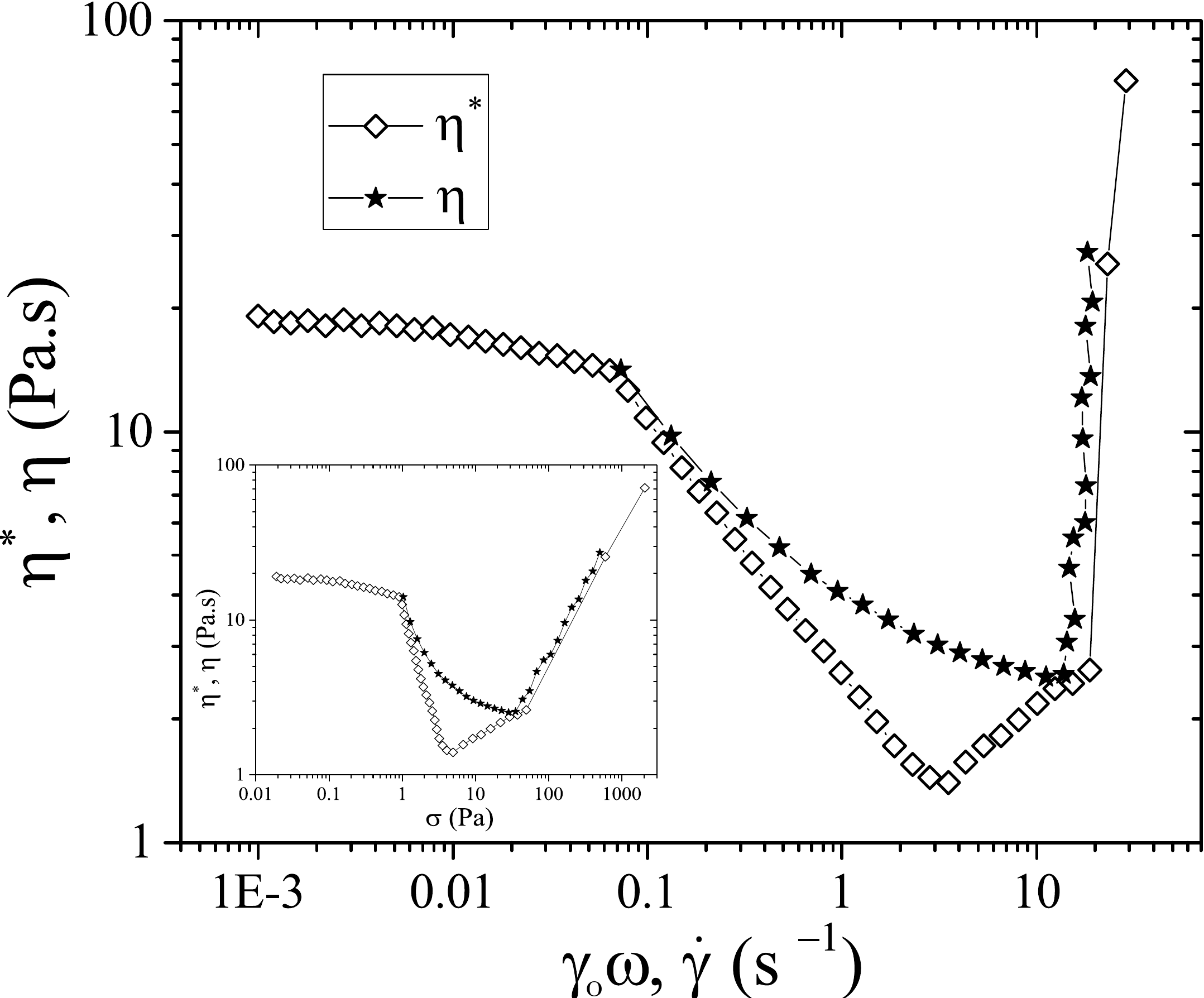}
\caption{ Steady shear flow curve showing   viscosity ($\eta$) vs shear rate ($\dot{\gamma}$, stars) and oscillatory shear showing complex viscosity ($\eta^{*}$) vs peak shear rate ($\gamma_o\omega$, where $\gamma _o$ is the strain amplitude and $\omega$ is the angular frequency, diamonds),   for volume fraction ($\phi$) = 0.56. The inset shows the same data plotted vs. shear stress (see Methods).}
\label{FC}

\end{figure}

The flow curve (viscosity measured during steady shear) for $\phi = 0.56$, shown in Fig. \ref{FC} (stars), exhibits the behavior typical of suspensions undergoing DST, with a region of shear thinning at low applied shear, followed by an abrupt increase in viscosity at a critical shear rate $\dot\gamma _c \sim$ 20 s$^{-1}$.  The open diamonds  in Fig. \ref{FC} show the complex viscosity $\eta ^*$ determined by applying an oscillatory strain  $\gamma(t) = \gamma_{o} \sin(\omega t)$ at fixed  $\omega = 1$ rad/sec and increasing $\gamma_{o}$.  The oscillatory rheology reveals  a two step shear thickening, where during first phase (peak shear rates $\sim 3 < \gamma_o \omega < \sim 20$ s$^{-1}$, corresponding to strain amplitudes $\sim 3 < \gamma_o < \sim 20$) the increase in the complex viscosity $\eta^{*}$ is gradual, until the peak shear rate approaches $\dot\gamma _c$ for DST, at which point $\eta ^*$ increases abruptly. Similar behavior was observed in LAOS of suspensions of monodisperse silica spheres by Lee et al, \cite{Lee:2018aa}, where simultaneous small angle x-ray scattering revealed that the transition to the first shear thickening regime is associated with a strain-controlled order-to-disorder transition that disrupts the ordering-induced shear thinning that is observed at smaller amplitudes.  This first shear thickening phase is not observed in steady shear, presumably because the oscillatory shear generates more substantial ordering, and therefore a lower viscosity, than steady shear.  The second, more abrupt thickening transition occurs in both steady shear and LAOS, at roughly the same shear rate (Fig. \ref{FC}) and peak shear stress.

\begin{figure*}
\includegraphics[width=0.8\textwidth]{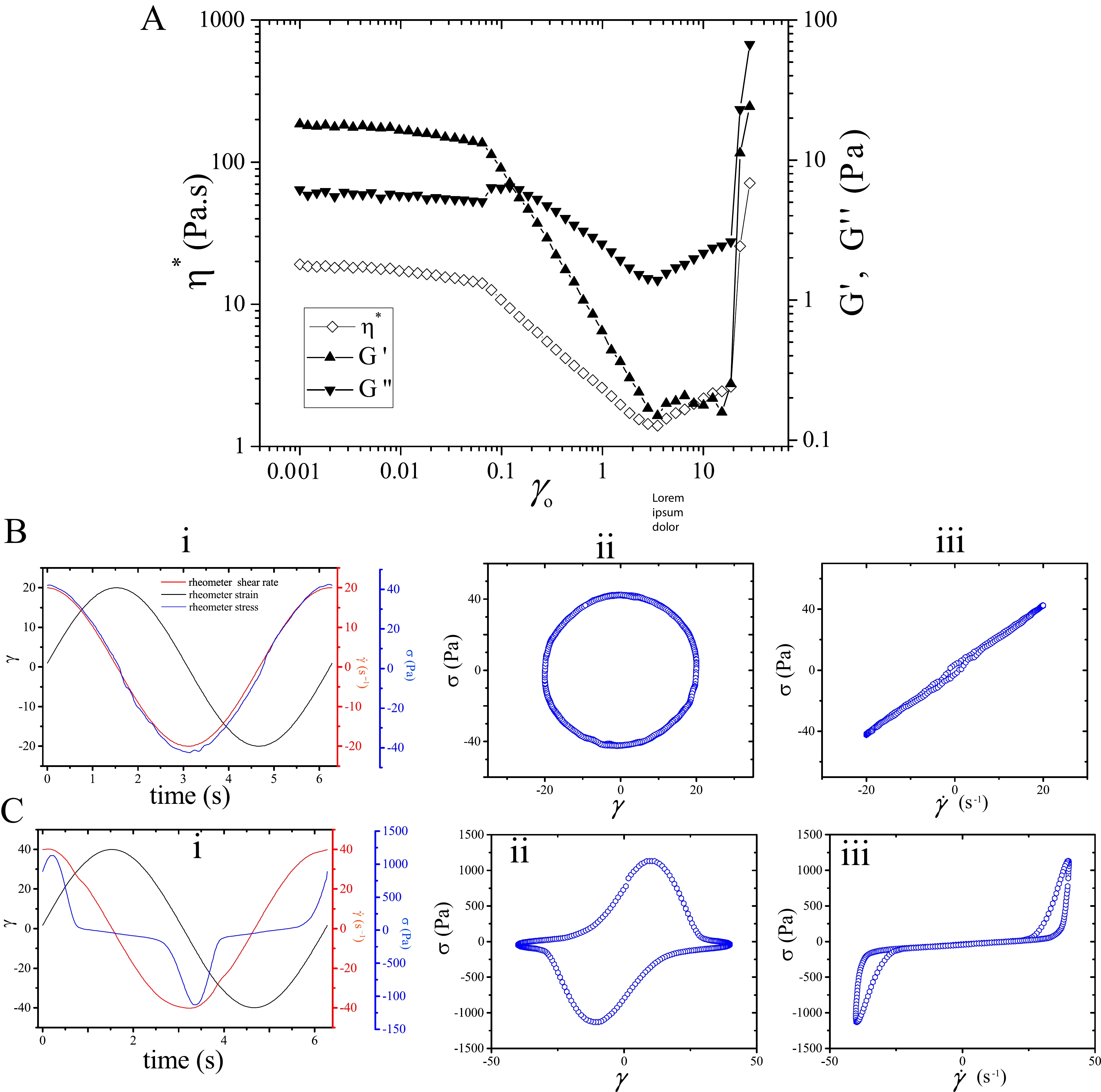}
\caption{ (A) Complex viscosity, $\eta ^*$, and elastic and loss modulii ($G'$ and $G''$, respectively) from an amplitude sweep  at fixed frequency ($\omega$ = 1 rad/s). (B-C):  $i$ Strain,  strain rate, and stress vs. time; $ii$ stress vs. strain  (Elastic Lissajous Figure); and $iii$ stress vs. strain rate (Viscous Lissajous Figure) at (B) $\gamma$  = 20 and (C)  $\gamma$  = 40.  } 
\label{LJ}
\end{figure*}

Figure \ref{LJ}A shows the complex viscosity data from Fig. \ref{FC} compared with $G'$ and $G''$ from the same measurement.  At low frequencies, the elastic response dominates the viscous, with a crossover for $\gamma >0.1$, and both components show the same shear thinning and two stage shear thickening behavior seen in the viscosity.  Note that the transition from the linear regime, with a  roughly constant modulus at very low amplitudes, to shear thinning at $\gamma \sim 0.1$ is reminiscent of a yield stress behavior.

The data in Figure  \ref{LJ}A include only the amplitude of the fundamental component of the stress response, and as such provide limited information about the nonlinear response.  Additional insight can be obtained from investigating the stress vs. time (Fig. \ref{LJ}B,C $i$) and the stress vs. strain (Fig. \ref{LJ}B,C $ii$) or strain rate (Fig. \ref{LJ}B,C $iii$), known as  Lissajous Figures. In the linear regime (Small Amplitude Oscillatory Shear, or SAOS), the stress response is harmonic, and because it is dominated the viscous component, closely tracks the sinusoidal applied strain rate (Fig. \ref{LJ}B$i$), with a linear, closed viscous Lissajous Figure  (Fig. \ref{LJ}B$iii$), indicating that the stress is everywhere nearly proportional to the strain rate.   The figures look roughly similar throughout the shear thinning and first shear thickening regimes ($\gamma < 20$, data not shown) because that the stress signal is primarily harmonic, indicating that the viscosity of the suspension is not changing during the oscillation cycle, even though it changes with oscillation amplitude.  This suggests that the oscillation-induced ordering that underlies the non-Newtonian behavior of the complex viscosity in this regime persists throughout the oscillation cycle.   

The situation changes dramatically at higher strain amplitudes, where the stress response becomes strongly non-harmonic.  This can be seen clearly in Fig. \ref{LJ}C$i$, which shows the stress vs time at $\omega=1$ and $\gamma=$40. As in earlier studies (\cite{Khandavalli:2015aa, Ness:2017aa}, a dramatic increase in stress is observed when the shear rate exceeds the value for DST in steady shear, as can be seen most clearly in the viscous Lissajous Figure ($\sigma$ vs. $\dot\gamma$, Fig. Fig. \ref{LJ}C$ii$).  As shown below, this abrupt increase in stress is a signature of the onset of large stress heterogeneities.

\section{Boundary stress microscopy}
\begin{figure}
	\includegraphics[width=0.45\textwidth]{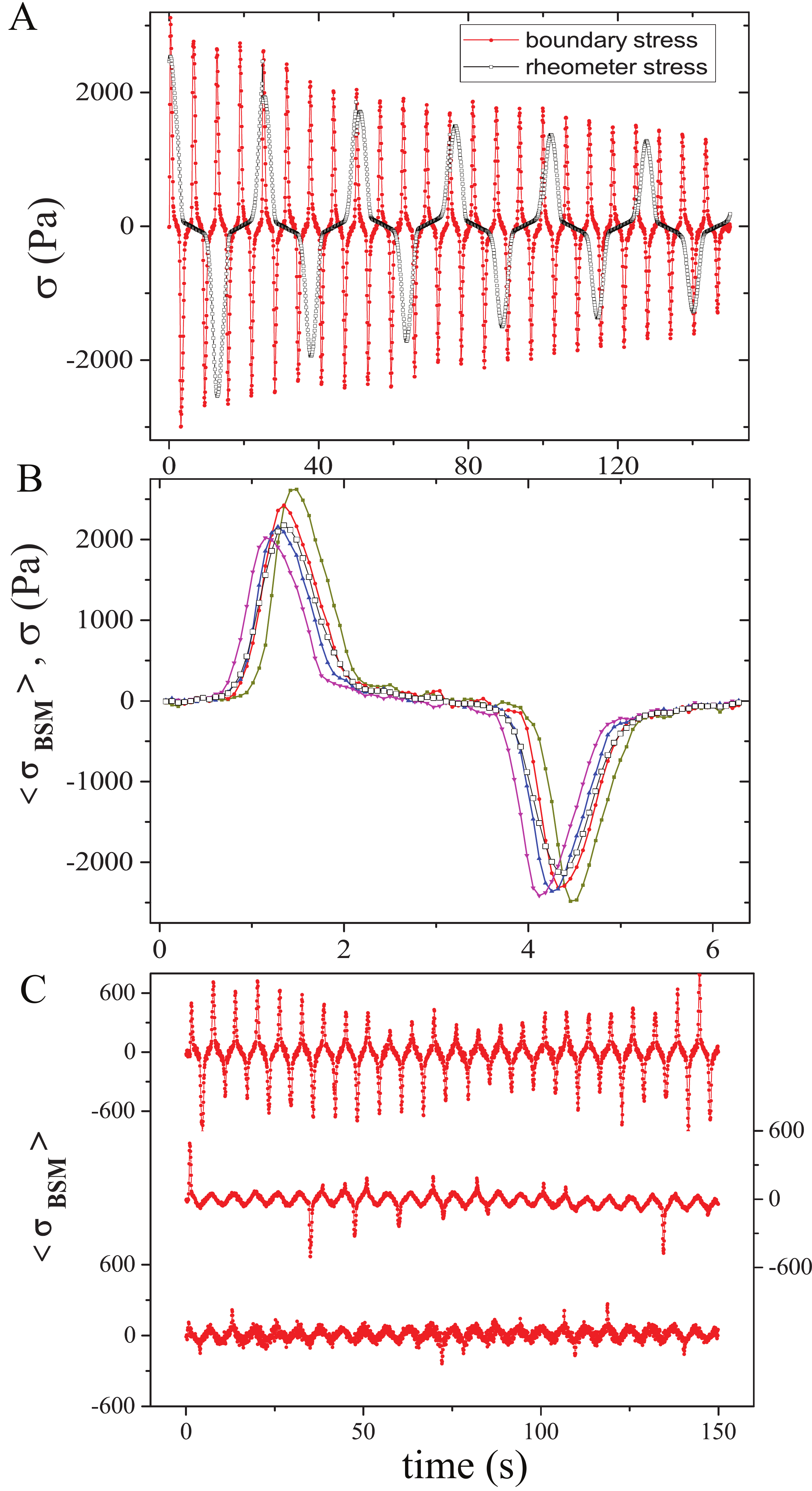}
	\caption{ A) Rheometer stress, $\sigma$, and $\sigma_{\rm{BSM}}$ vs time  at $\gamma _o$  = 50 and $\omega$ = 1 rad/s.    B) $\sigma_{\rm{BSM}}$ from four consecutive cycles, and the rheometer stress reported for those four cycles (thick line with open circles).
	(C) $\sigma_{\rm{BSM}}$ vs. time at  $\gamma _o$ = 26 (bottom), 30 (middle) and 40 (top) at $\omega$ = 1.
	} 
	\label{BSM_avg}
\end{figure}

Boundary Stress Microscopy provides a direct readout of the spatially resolved stress at the bottom boundary of the suspension.  We have shown that the net torque calculated from this stress matches the stress inferred from the torque on the rheometer tool (the reported rheomter stress) during steady shear when the entire sample surface is imaged in BSM \cite{Rathee:2020aa}.  A direct comparison is complicated in this case in part because only a portion of the sample is imaged (see Methods), but also because in LAOS mode the Anton-Paar rheometer reports stress vs time averaged over 4 cycles.  In  Fig. \ref{BSM_avg}A we directly compare the average stress in each BSM image, $<\sigma_{\rm{BSM}}>$, with the stress vs. time from the rheometer.    Unlike the rheometer output, BSM provides independent time-resolved measurements for each cycle.  Note that both the rheometer and BSM show a gradual reduction in peak stress over the course of the measurement.  Slow drifts like this are often seen in long experiments and can arise from a variety of sources including particle settling or migration, or slight mixing of the suspension and the surrounding oil at the boundary of the sample. 

Figure \ref{BSM_avg}B shows $<\sigma_{\rm{BSM}}>$ for four consecutive cycles, compared to the stress vs. time reported by the rheometer for those same cycles.   The data acquisition for BSM is not synchronized with the rheometer, so the timing for each cycle is approximate, but overall $<\sigma_{\rm{BSM}}>$ matches the rheometer stress quite well.  Fig.  \ref{BSM_avg}C shows the evolution of the dramatic shear thickening, from appearing only on a few cycles at $\gamma _o =26$ to appearing on every cycle at $\gamma _o=40$

\begin{figure*}
\includegraphics[width=1\textwidth]{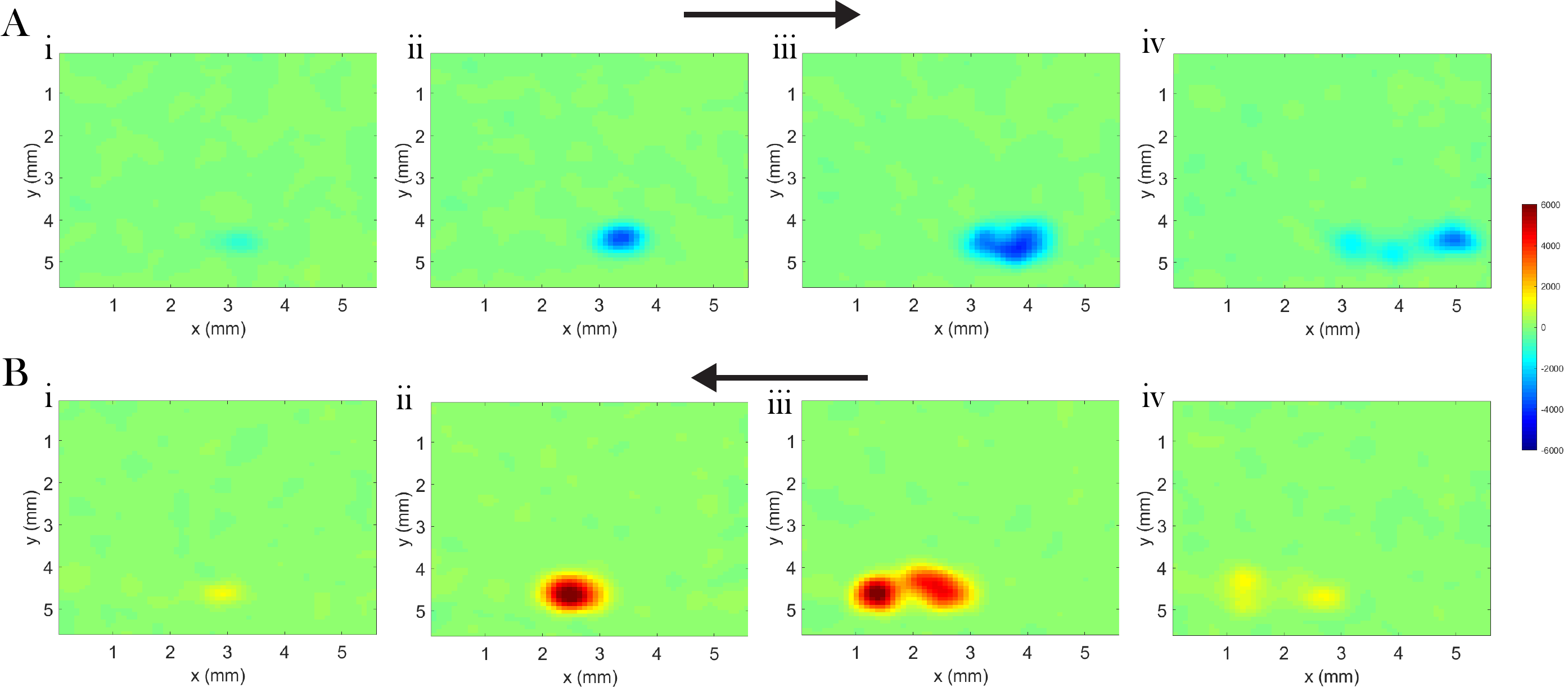}
\caption{Spatiotemporal dynamics of boundary stress during one complete cycle at $\gamma _o$ = 26 and $\omega$ = 1 rad/s. A (B) Sequence of images corresponding to increasing and then decreasing positive (negative) shear rate. The arrows show direction of shear. } 
\label{stress_map_26}
\end{figure*}

The real power of BSM, however, lies in its ability to reveal spatiotemporal dynamics of boundary stresses.   Figure \ref{stress_map_26} shows a series of images of the spatial distribution of boundary stresses during one cycle of the run shown in Fig. \ref{BSM_avg}C for $\gamma=26$ (see also supplementary movie SM1 \cite{SM_movie_link}). During the phase of the cycle where the shear rate is low (A,B $i$), the boundary stress is low and spatially uniform, to within the resolution of the measurement.  When the shear rate gets large enough, however, localized regions of high stress appear ($ii$) and then bifurcate ($iii$), and finally disappear as the shear rate drops ($iv$).  It is clear from the time evolution (see SM1) that one of the bifurcated regions stays fixed, while the other region propagates in the shear direction.   The behavior is strikingly similar to the dynamics we observed under steady shear close to the critical shear rate, where BSM revealed the localized transitions to fully jammed solid-like phases (SLPs) \cite{Rathee:2020aa}.  After formation, the SLP rapidly fractures, producing two separate SLPs, one that is nearly fixed in space and presumably in contact with the bottom plate, and the other that appears to be moving with the top plate.  We surmise that the same dynamic occurs in LAOS, on each cycle when the shear rate exceeds the critical shear rate for DST. We note that the size of the high stress regions is on the order of a few hundred microns, much larger than the particle size but of the same order of magnitude as the gap between the rheometer plates ($\sim 125 \mu$m where the BSM is measured).

\begin{figure*}
\includegraphics[width=1\textwidth]{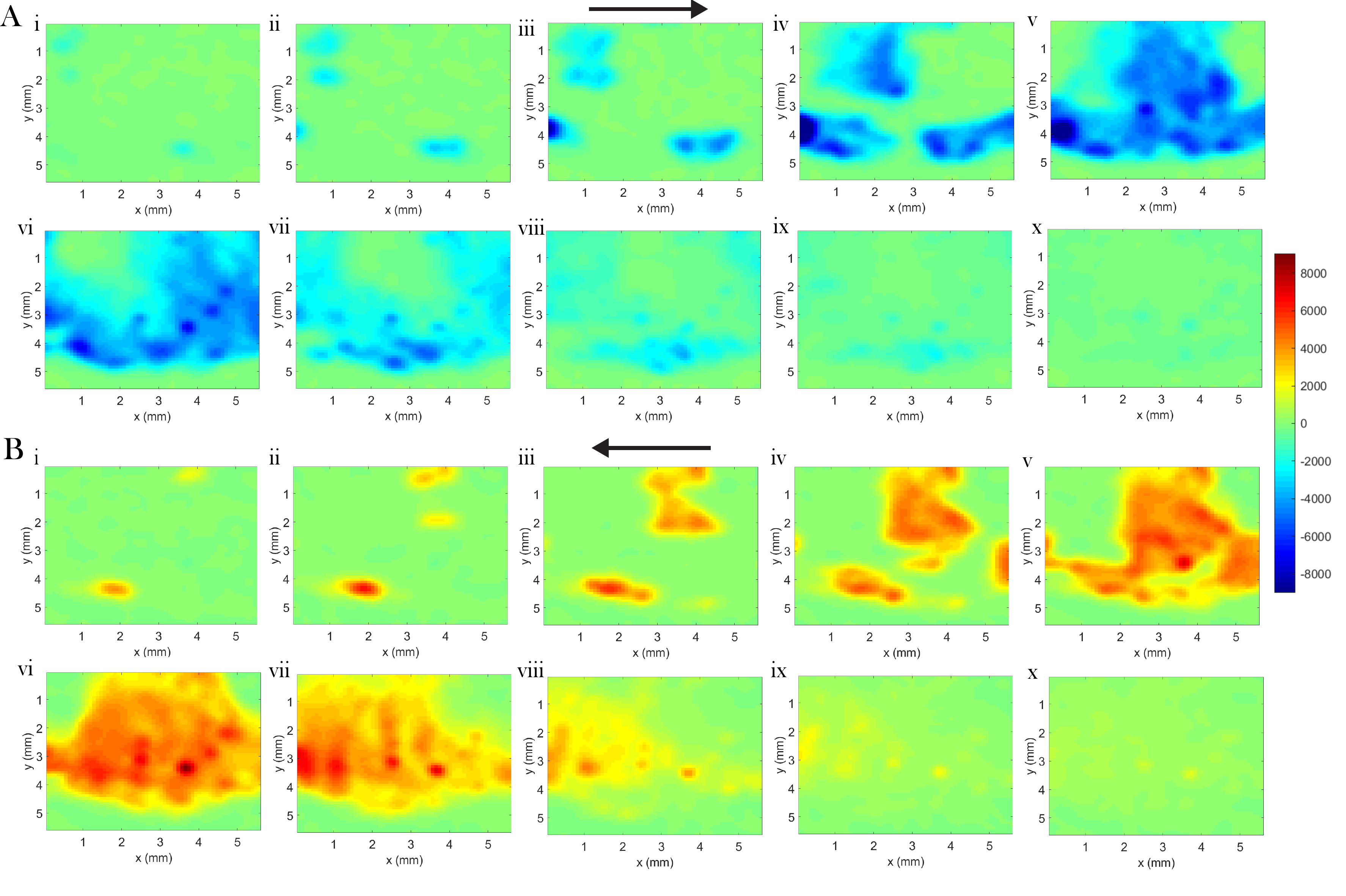}
\caption{ Spatiotemporal dynamics of a high stress event during one complete cycle at $\gamma$ = 50 and $\omega$ = 1 rad/s.  (A) and  (B) Sequence of images corresponding to increasing and then decreasing positive (negative) shear rate. The arrows show direction of shear. } 
\label{stress_map_50}
\end{figure*}

At higher strain amplitudes,  the high stress regions cover more of the sample boundary and live longer, presumably because the system spends more time above $\dot\gamma _c$. Figure \ref{stress_map_50} shows a series of images taken during one cycle of a measurement with $\gamma=50$ and $\omega$ = 1 (see also supplementary movie SM2). During the phases of the cycle where the shear rate is low (Fig. \ref{stress_map_50} A$i,ix$, B$x$) the boundary stress is uniform and low.  As the shear rate increases, localized regions of high stress appear and  bifurcate (Fig. \ref{stress_map_50} A$iii$, B$i,ii$), as at lower strain amplitudes.  As the shear rate increases, these high stress regions grow in extent and new regions appear (Fig. \ref{stress_map_50} A$iv-vi$, B$ii-vii$), and then finally shrink and disappear as the shear rate decreases  (Fig. \ref{stress_map_50} A$vii-ix$, B$viii-ix$).

\begin{figure}[H]
\includegraphics[width=0.45\textwidth]{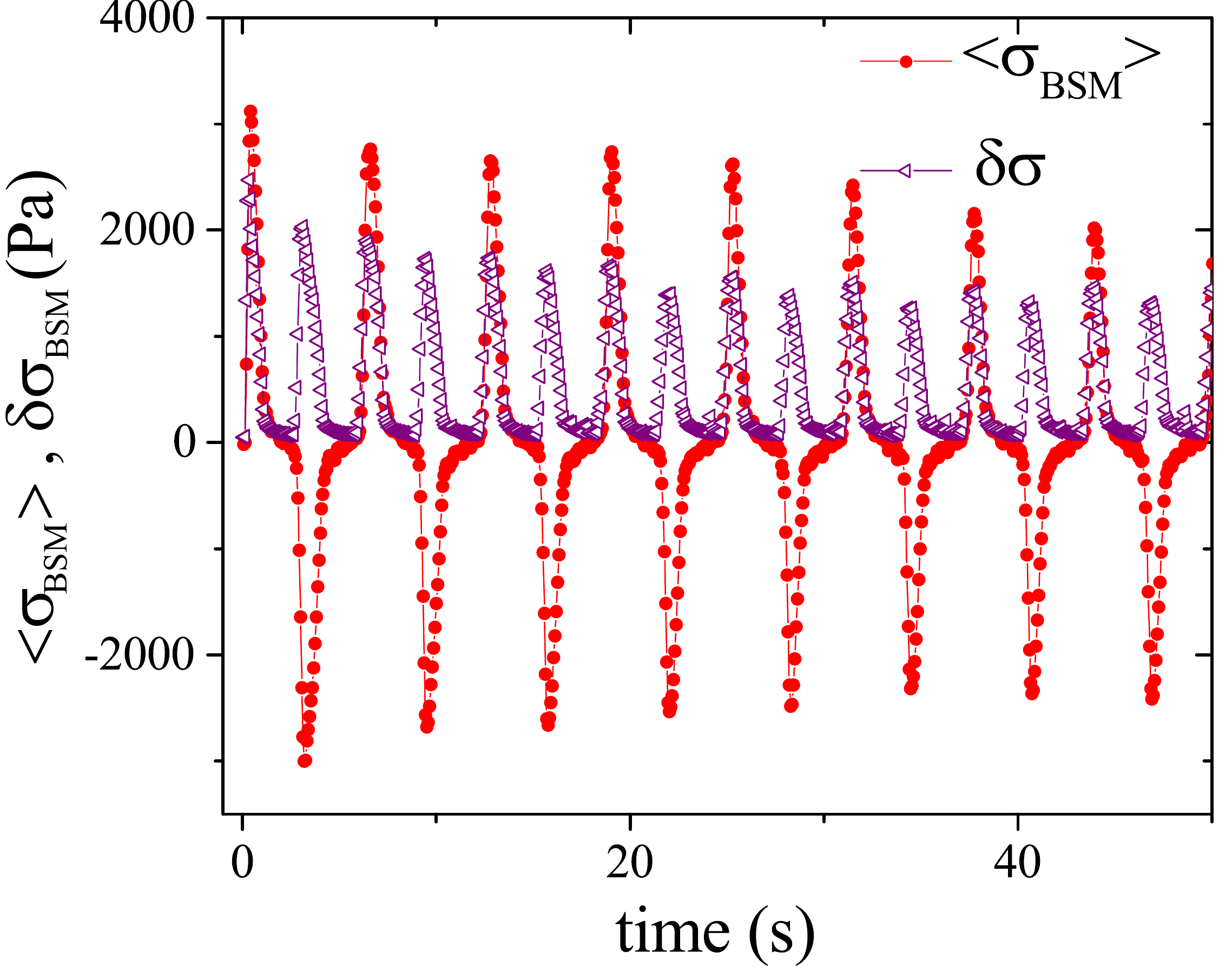}
\caption{Average boundary stress $<\sigma _{\rm{BSM}}>$ and stress roughness $\delta \sigma _{\rm{rms}}$ vs. time for  $\gamma _o$ = 50 and $\omega$ = 1 rad/s.   } 
\label{roughness}
\end{figure}

The spatial heterogeneity or roughness of the boundary stress can be quantified by calculating the root mean square stress deviation in each image,  $\delta \sigma _{\rm{rms}}$, calculated by subtracting $<\sigma _{\rm{BSM}}>$ from each pixel, and computing the rms of the resulting stress variation for each frame.
Figure \ref{roughness} shows a plot of $<\sigma _{BSM}>$ and $\delta \sigma _{\rm{rms}}$ vs. time for  $\gamma$ = 50 and $\omega$ = 1 rad/s. For large stresses, where $\dot\gamma$ exceeds $\dot\gamma _c$, the roughness closely tracks the absolute value of the mean stress,  while it remains close to zero when the mean stress is small. Thus the roughness provides a very clean measure of the presence of local high viscosity or jammed phases.  It is evident from Fig. \ref{roughness} that the roughness grows rapidly but decays slowly, similar to our previous measurements of high stress fluctuations at constant shear rate \cite{Rathee:2020aa}. 


\begin{figure*}
\includegraphics[width=1\textwidth]{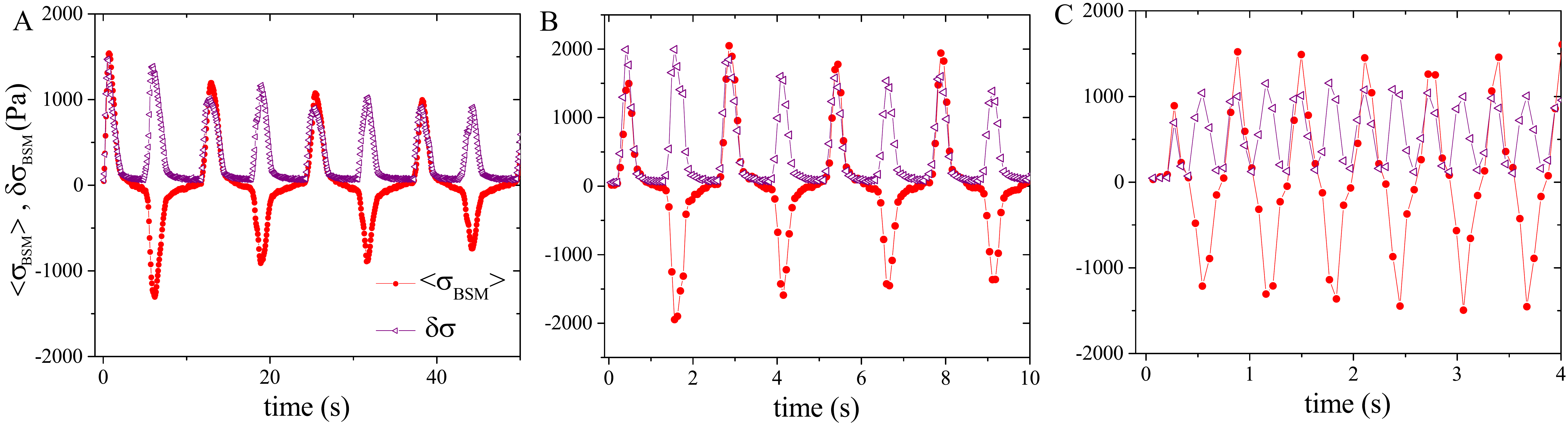}
\caption{Average boundary stress $<\sigma _{BSM}>$ and stress roughness $\delta \sigma _{\rm{rms}}$ vs. time for  (A) $\gamma _o$ = 80 and $\omega = 0.5$ rad/s, (B) $\gamma _o$ = 20 and $\omega$ = 2.5 rad/s, and (C) $\gamma _o$ = 9 and $\omega$ = 10 rad/s.   } 
\label{roughness_omega}
\end{figure*}

\begin{figure}
\includegraphics[width=0.45\textwidth]{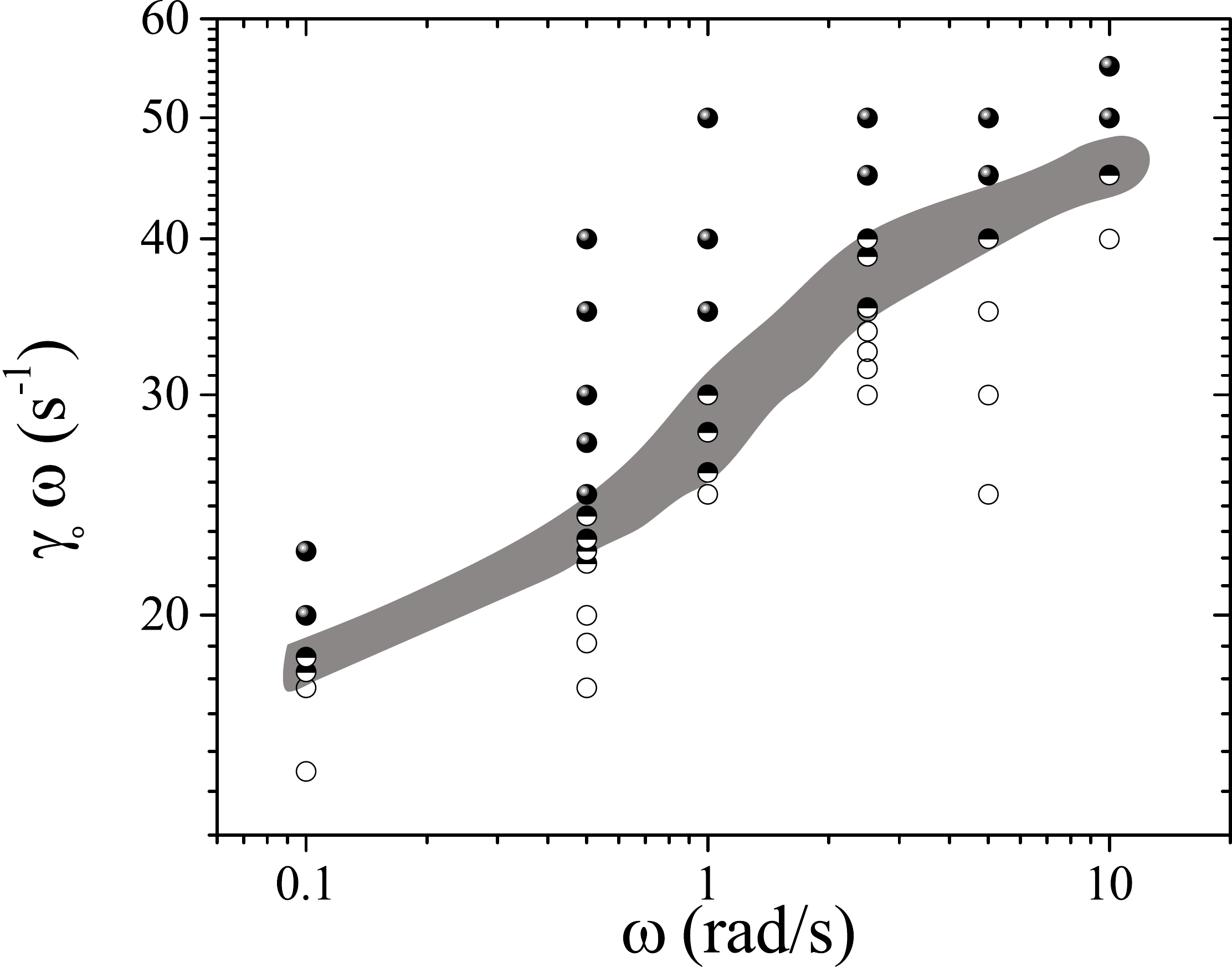}
\caption{ Pipkin diagram of peak shear rate, $\gamma _o \omega$, vs frequency ($\omega$), showing regions without heterogeneous stresses (open circles), regions that show them on some but not all cycles (half filled circles), and those that show them on every cycle (filled circles) at $\phi$ = 0.56.  } 
\label{Pipkin}
\end{figure}

We observe a similar behavior at higher $\omega$, with heterogeneous high stresses appearing when $\gamma \omega > \gamma_c$. Figure \ref{roughness_omega} shows  representative roughness data  at $\omega = 0.5$ (A), 2.5 (B), and 10 (C) rad/sec, at strain amplitudes such that the peak shear rate is large enough to produce heterogeneous stresses on every cycle.  At all frequencies the regions of large heterogeneous stresses are separated by periods when the stress is small and uniform.  Those periods are barely resolvable at the $\omega$ = 10 rad/s (Fig. \ref{roughness_omega}C), because our imaging speed (28 frames/sec) provides only about 5 frames per cycle.  Note that our imaging is performed by laser scanning confocal microscopy, and the images shown are acquired from top to bottom, with acquisition of the bottom of one frame almost immediately followed by acquisition of the top of the next frame.  Careful inspection of the movie from $\omega$ = 10 rad/s (Movie SM3) reveals a noticeable phase lag between the top and bottom the frames, but every cycle shows some regions with uniform low stress.

While the evolution from of the stress field from uniform low stress to heterogeneous high stress does not change as the frequency is increased, we do observe two notable changes.  One is that peak shear rate required to generate heterogeneous high stresses increases with frequency.  Figure  \ref{Pipkin} shows a Pipkin diagram, with open circles marking conditions that show only uniform low stress, filled circles indicate conditions where significant roughness is measured on every cycle, and the half-filled circles are conditions that produced high stress regions on some but not all cycles (e.g the regime shown in Fig. \ref{BSM_avg}C, $\sim 25 < \gamma _o < 40, \omega = 1$ rad/sec ).  The crossover region (shading in Fig. \ref{Pipkin}) increases from $\approx 17 $ sec$^{-1}$ at $\omega$ =0.1 (approximately equal to $\gamma _c$ measured by steady shear) to $\approx 45 $ sec$^{-1}$ at $\omega$ = 10 rad/s.  Note that the strain amplitude in the crossover region decreases from $\gamma_{o}\approx$ 170 at $\omega$ =0.1 rad/s to $\gamma_{o}\approx$ 4.5 at $\omega$ = 10 rad/s.    As discussed below, the frequency dependence of the shear rate required to initiate heterogeneous stresses is consistent with a frequency independent critical shear stress, and a frequency dependence to the viscosity before thickening is observed.

\begin{figure*} 
\includegraphics[width=1\textwidth]{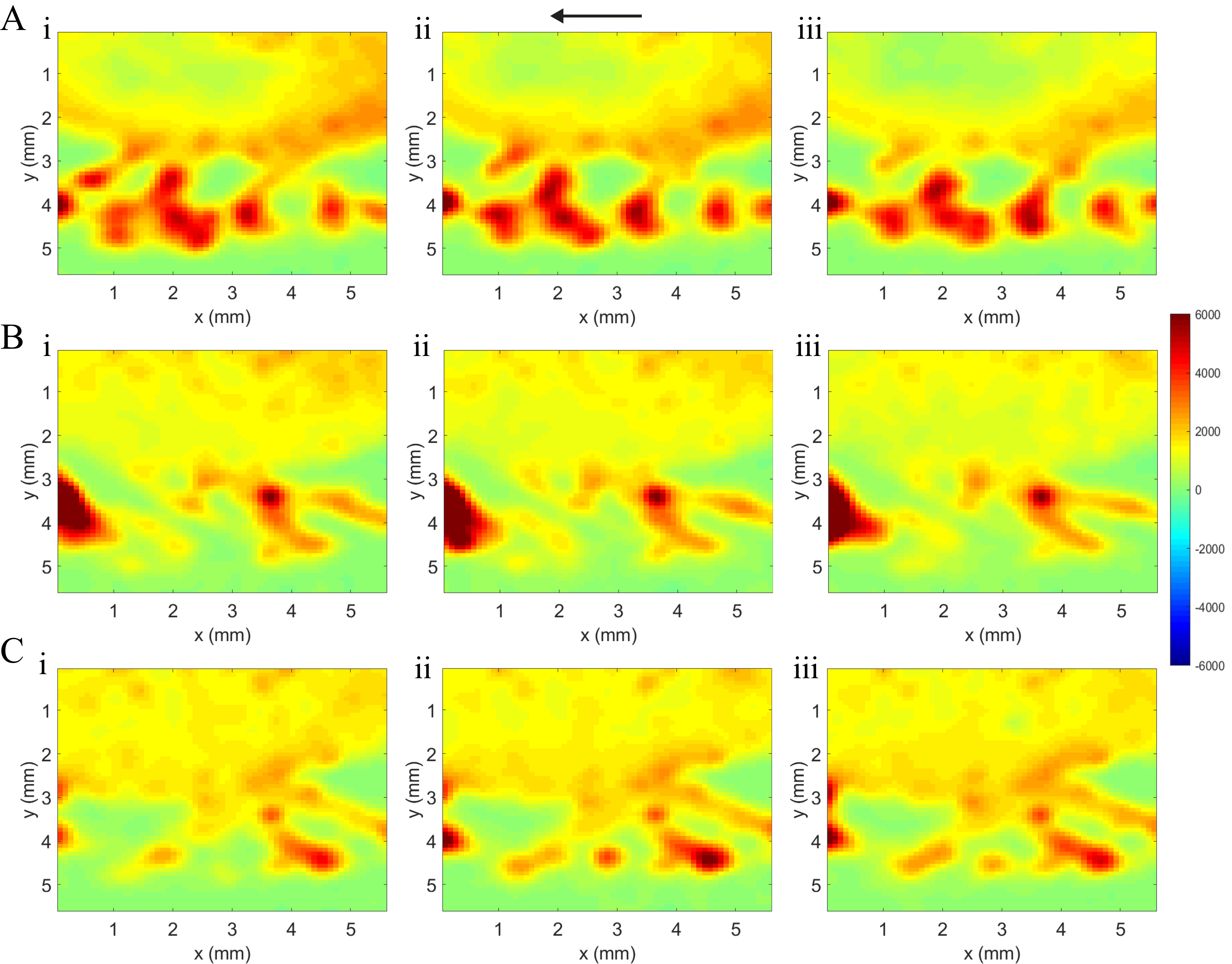}
\caption{Pattern of boundary stresses at the peak positive shear rate on three consecutive cycles at the beginning (A), middle (B), and end (C) of a LAOS measurement at $\gamma _o$ = 9 and $\omega$ = 10 rad/s. There are approximately 50 cycles between (A) and (B) and between (B) and (C). Also see supplemental movie SM3 \cite{SM_movie_link}. The arrow shows direction of shear.} 
\label{stress_map_omega10}
\end{figure*}

The other notable change in the spatiotemporal dynamics of the boundary stresses as frequency increases is an increase in the persistence of the pattern of stresses on successive cycles. This is visually evident when comparing the movies from $\omega = 1$ and 10 rad/s (SM2 and SM3, respectively \cite{SM_movie_link}), both with peak shear rates substantially above $\dot\gamma _c$. Figure \ref{stress_map_omega10}A shows the pattern of boundary stresses measured near the peak shear rate on three consecutive cycles at the beginning of a LAOS measurement lasting approximately 100 cycles (from movie SM3 \cite{SM_movie_link}). The pattern is disordered, but the change from one cycle to the next is quite small, despite the fact that the cycles are separated by periods when the shear rate is low and no high stresses are visible.  Figure \ref{stress_map_omega10}B and \ref{stress_map_omega10}C show stress patterns from three consecutive cycles at the middle and end, respectively, of the LAOS run.  The patterns evolve over the course of many cycles, but consistently show a high degree of persistence between cycles.   The implications of this observation are discussed below.



\section{Discussion and Conclusions}
\label{discussion}
The results reported here are generally consistent with previous investigations of shear thickening colloidal suspensions during LAOS  \cite{Khandavalli:2015aa,Lee:2018aa} and our observations of boundary stresses in suspensions undergoing DST during steady shear \cite{Rathee:2020aa}:  on each cycle, upon increasing shear rate, if and when the average stress exceeds a critical stress $\sigma _c$, large heterogenous boundary stresses appear  in a manner consistent with the localized formation of fully jammed gap spanning solid like phases (SLPs) with a spatial extent in the flow-vorticity plane comparable to the rheometer gap.  The imposed strain produces a rapid fracture of the SLPs, with pieces staying adhered to the top (moving) and bottom (stationary) boundaries.  The SLPs and the associated large boundary stresses persist until the stress falls  sufficiently below  $\sigma _c$ as a result of the decreasing imposed shear rate.  As in steady shear, the onset of the large stress heterogeneity on each cycle is nearly instantaneous, while the decay is relatively gradual. 

The frequency dependence of the minimum strain amplitude necessary to generate large stresses  rules out a strong role for strain amplitude.  For example, at $\omega =10 $ rad/s, the onset occurs at a strain $\gamma _o \sim 4$, whi1e it occurs at $\gamma _o \sim 25$ at $\omega =1$ rad/s and $\gamma _o >  150$ at $\omega =1 $ rad/s (Fig.  \ref{Pipkin}).  The behavior of dense suspensions under oscillatory shear is sensitive to strain amplitudes for $\gamma_o \lesssim$ 1 \cite{Ackerson:1990aa, Ness:2017aa, Galloway:2020aa}, however for the frequencies studied here strain amplitudes substantially larger than unity are required to reach $\sigma _c$.  Similarly, the weak frequency dependence shown in Fig. \ref{Pipkin} suggests that the Peclet number associated with the oscillatory period, $Pe_\omega = D \omega / R ^2$, where $D$ is the particle diffusion coefficient and $R$ the radius, plays a minor role.  Using the bare diffusion coefficient for the $0.75 \mu$m radius particles suspended in 80:20 gycerol:water used here, $Pe_\omega$ ranges from $\sim 10$   at $\omega = 0.1$ rad/s to $\sim 1000$ at $\omega = 10$.  Thus in the regimes studied here, DST  always occurs at either very high $Pe_ \omega$ or very high strain amplitude, so the  regimes where Brownian motion has a significant impact on dense suspensions during LAOS, for example by enabling cage escape by Brownian motion \cite{Koumakis:2013ab,Koumakis:2016aa}, are not encountered.

Consistent with previous results \cite{Lee:2018aa}, the transition to DST in LAOS for the suspension studied here is likely determined by $\sigma _c$, so the frequency dependence observed in Fig. \ref{Pipkin}  arises from a modest decrease in suspension viscosity as the oscillation frequency is increased, attributable to increased ordering of the nearly monodisperse particles \cite{Lee:2018aa}.  Shear-induced ordering is common in dense monodisperse suspensions undergoing LAOS \cite{Ackerson:1990aa,Besseling_2012}, and can be quite complex, particularly for small amplitude oscillations.   Ordering, both in the gradient direction (layering) and within the layers that form in the flow-vorticity plane, is sensitive to the boundary conditions, and a solid planar surface like that at the base of our rheometer can provide a nucleation site for crystals \cite{Shereda:2010aa}, and stabilize substantial layering many particle diameters away from the planar surface \cite{Ramaswamy:2017aa, Pieper:2016aa}.  The layering reduces the suspension viscosity, because the particles can flow past each other with reduced particle interactions, so the increased layering near the boundary will produce a local decrease in viscosity \cite{Zhao:2016aa, Gallier:2016aa}, and a nonuniform shear profile.   It is well established that bulk order-disorder transitions are not required for DST (reviewed in \cite{Morris:2020aa}), and in particular it was shown in \cite{Lee:2018aa} that a small amount of polydispersity greatly reduces bulk ordering but doesn't affect the rheology at and above $\sigma _c$.  Layering that is confined to a region close to the suspension boundary, however, will not be detectable by bulk probes like the SAXS used in \cite{Lee:2018aa}, but may be robust to polydispersity and have a substantial impact on the dynamics of the instabilities that are observed during DST. 

A striking feature revealed by BSM is the remarkable persistence of the stress patterns observed at high frequency (Fig. \ref{stress_map_omega10} and movie SM3 \cite{SM_movie_link}), despite strain amplitudes $\sim 10$, much larger than those that produce reversible trajectories \cite{Ness:2017aa, Galloway:2020aa} or hyperuniform structures  \cite{Wilken:2020aa} in non-Brownian suspensions.  A recent study using shear reversal during LAOS revealed that residual structural anisotropy in the particle contact network provides a source of memory in dense suspensions \cite{Ong:2020aa}, but the patterns observed here are large compared to the size of the individual particles.  We note that the SLPs observed in steady shear persist for very large strains ($>$ 100) \cite{Rathee:2020aa}.  We hypothesize that both the stability of SLPs in steady shear and the pattern persistence in LAOS arise from the presence of significant spatial variations in concentration,  layering or other ordering over relatively large spatial scales.

In summary, the results presented hear show that dense suspensions that undergo discontinuous shear thickening during LAOS exhibit large spatial heterogeneous boundary stresses indicative of the cyclic formation and breakup of fully jammed solid-like phases.   At high frequencies, memory of these heterogeneities  persists despite large strain amplitudes.  Uncovering the source of this memory will likely require measurements that can track spatial fluctuations in particle density, layering, and order during the onset of shear thickening.  

\section{Acknowledgements} The authors thank Emanuela Del Gado and Peter Olmsted for helpful discussions. This work was supported by the National Science Foundation (NSF) under Grant No. DMR-1809890. J.S.U. is supported, in part, by the Georgetown Interdisciplinary Chair in Science Fund.

\bibliography{LAOS_BSM_bib}

\clearpage
\newpage

\begin{center}
    \section*{Supplementary Material}
\end{center}

\section*{Supplementary Movie Captions}
\noindent\textbf{Supplementary Movie 1:} Spatiotemporal dynamics of high stress events revealed by boundary stress measurements at frequency ($\omega$) = 1 and strain ($\gamma_{o}$)  = 26,  for $\phi$ = 0.56. The total time elapsed is  150s having total 24 cycles. The corresponding snapshots are shown in Figure \ref{stress_map_26}. The movie is played at 15 fps.

\noindent\textbf{Supplementary Movie 2:} Spatiotemporal dynamics of high stress events revealed by boundary stress measurements at frequency ($\omega$) = 1 and strain ($\gamma_{o}$)  = 50,    for $\phi$ = 0.56. The total time elapsed is 150s having total 24 cycles. The corresponding snapshots are shown in Figure \ref{stress_map_50}. The movie is played at 15 fps.

\noindent\textbf{Supplementary Movie 3:} Spatiotemporal dynamics of high stress events revealed by boundary stress measurements at frequency ($\omega$) = 10 and strain ($\gamma_{o}$)  = 9,  for $\phi$ = 0.56. The total time elapsed is 90s having total 144 cycles. The corresponding snapshots are shown in Figure \ref{stress_map_omega10}. The movie is played at 9 fps.

\end{document}